\documentclass[a4paper,11pt]{article}
\usepackage{pos}
\usepackage{tikz}
\usepackage{pgfplots}
\usepackage{graphicx}
\usepackage{caption}
\usepackage{subcaption}

\usetikzlibrary{arrows.meta}
\usetikzlibrary{backgrounds}
\usepgfplotslibrary{patchplots}
\usepgfplotslibrary{fillbetween}
\pgfplotsset{%
    layers/standard/.define layer set={%
        background,axis background,axis grid,axis ticks,axis lines,axis tick labels,pre main,main,axis descriptions,axis foreground%
    }{
        grid style={/pgfplots/on layer=axis grid},%
        tick style={/pgfplots/on layer=axis ticks},%
        axis line style={/pgfplots/on layer=axis lines},%
        label style={/pgfplots/on layer=axis descriptions},%
        legend style={/pgfplots/on layer=axis descriptions},%
        title style={/pgfplots/on layer=axis descriptions},%
        colorbar style={/pgfplots/on layer=axis descriptions},%
        ticklabel style={/pgfplots/on layer=axis tick labels},%
        axis background@ style={/pgfplots/on layer=axis background},%
        3d box foreground style={/pgfplots/on layer=axis foreground},%
    },
}

\newcommand{\todo}[1]{}
\renewcommand{\todo}[1]{{\color{red} TODO: {#1}}}
\newcommand{\csw}{c_{\mathrm{SW}}}
\newcommand{\mps}{m_{\mathrm{PS}}}
\newcommand{\mv}{m_{\mathrm{V}}}
\title{$2$-flavour $SU(2)$ gauge theory with exponential clover Wilson fermions}

\author*[a]{Laurence Sebastian Bowes}
\author[a]{Vincent Drach}
\author[b]{Patrick Fritzsch}
\author[a,c]{Antonio Rago}
\author[d]{Fernando Romero-Lopez}

\affiliation[a]{Centre for Mathematical Sciences, University of Plymouth\\England, UK}
\affiliation[b]{School of Mathematics, Trinity College Dublin\\Ireland}
\affiliation[c]{IMADA and Quantum Theory Center, University of Southern Denmark\\Denmark}
\affiliation[d]{Center for Theoretical Physics, Massachusetts Institute of Technology\\USA}
\emailAdd{laurence.bowes@plymouth.ac.uk}
\emailAdd{vincent.drach@plymouth.ac.uk}
\emailAdd{fritzscp@tcd.ie}
\emailAdd{rago@qtc.sdu.dk}
\emailAdd{fromerol@mit.edu}

\abstract{
Composite Higgs models are a class of models proposed to address the hierarchy and naturalness problems associated with the Standard Model fundamental scalar Higgs. $SU(2)$ with two fundamental flavours is a minimal model for the composite Higgs sector which is not yet ruled out by experimental data. We present lattice results for $SU(2)$ with two fundamental mass degenerate flavours. For the fermion action we use the new exponential clover Wilson fermion action, which offers $O(a)$ improvement. We discuss tuning the $\csw$ parameter through Schr\"{o}dinger functional simulations, the scale setting of the ensembles using the Wilson gauge flow, and the low energy spectroscopy of the theory including the masses of the pseudoscalar isotriplet Goldstone bosons and the vector isotriplet.
}
\FullConference{The 40th International Symposium on Lattice Field Theory (Lattice 2023)\\
July 31st - August 4th, 2023\\
Fermi National Accelerator Laboratory\\}

\begin{document}
\maketitle

\section{Introduction}
The Standard Model of Particle Physics has reigned supreme for 50 years as humanity's finest scientific achievement. Despite this, we know that it is certainly not the fundamental theory of nature. There are a number of tantalising directions of tension, among these the naturalness problem and the hierarchy problem. Models of particle physics where the Higgs boson is composite are proposed in order to address these problems~\cite{cacciapaglia2020fundamental}.

Composite Higgs models are realised by introducing a new confining strong sector analogous to QCD into the SM, providing a dynamical origin for the spontaneous breaking of electroweak symmetry. In this case, the physical Higgs state emerges either as a pseudo-Nambu-Goldstone boson or as a light scalar resonance (technicolour). These two scenarios are not mutually exclusive, and they are two extremes of a one-parameter space of theories parameterised by the vacuum misalignment angle $\theta$.

If there truly is a composite Higgs sector, then it will affect scattering processes currently being tested at the LHC\@. The long term goal for our research program is therefore to understand clearly how an undetected composite Higgs sector would affect scattering amplitudes and the observable Higgs boson phenomenology at the LHC.

A minimal model for a composite Higgs sector based on $SU(2)$ with two mass degenerate fundamental fermions has been introduced in~\cite{Cacciapaglia:2014uja}, and it has been previously studied on the lattice~\cite{Lewis_2012, Arthur_2016, arthur2016su2, Pica_2017, Drach_2018, janowski2019resonance}. The new strong sector in isolation features an enhanced $SU(4)$ flavour symmetry due to the pseudoreality of the $SU(2)$ representation, which is expected to spontaneously break down to $Sp(4)$. Therefore, the chiral symmetry breaking pattern is associated with 5 Goldstone bosons. However, we will classify states by isospin.
In the continuum the Lagrangian density of the strong sector in isolation is given by

\begin{equation}
\mathcal{L} = -\frac{1}{4} F_{\mu\nu}^{a}F_a^{\mu\nu} +  \overline{\mathfrak{u}}(i \gamma^\mu D_\mu - m)\mathfrak{u}  + \overline{\mathfrak{d}}(i \gamma^\mu D_\mu - m)\mathfrak{d},
\end{equation}
where $\mathfrak{u}$ and $\mathfrak{d}$ are Dirac fields.

With regards to our previously stated long-term goal of understanding scattering and phenomenology of composite Higgs models, the scattering properties of the isosinglet Lorentz scalar state (the $\sigma$) are of considerable interest. Previous work studying the composite Higgs sector in isolation from the Standard Model using an untuned ($\csw = 1$) Wilson clover action has shown the singlet state to be stable down to  $\frac{m_{\mathrm{V}}}{m_{\mathrm{{PS}}}} < 2.5$ towards the chiral limit under the new interaction~\cite{Drach_2022}, which is the current threshold we are aiming to move beyond. However, in that previous setup there were strong $O(a)$ effects for small lattice spacing, one of the facts which contributed to moving to the new setup with the exponential clover action used in this work.

In this study we present our setup and the tuning necessary to achieve $O(a)$ improvement, our scale setting strategy and initial spectroscopy results. We begin by explaining our lattice setup for the physics simulations and the Schr\"{o}dinger functional simulations required to tune the action. We then give more detail about the specifics of how the action is tuned, and go on to present the results of the physics simulations and some remarks on them.

\section{Lattice Setup}

Let us begin with discussing the lattice setup for investigating physical predictions of the composite Higgs sector in isolation. For the gauge part of the action we use the Wilson plaquette action, and for the fermion fields we employ the recently introduced Symanzik-improved exponential clover fermion action~\cite{Francis:2019muy}.  
The exponential clover action provides $O(a)$ improvement once the parameter $\csw$ is tuned non-perturbatively. Unlike the Wilson clover action, the exponential clover action also offers enhanced numerical stability by protecting against zero modes of the odd-odd block of the even-odd preconditioned Dirac operator. The non-perturbative tuning requires dedicated simulations that will be discussed below. The action we use can be written as
\begin{equation}
  S = \sum_x\left[\frac{\beta}{2} \sum_{\mu < \nu} \mathfrak{Re}\ \mathrm{Tr} [1 - P_{\mu, \nu}(x)] + \overline{\psi}(x)D\psi(x) \right],
\end{equation}
where the diagonal part of the even-odd preconditioned Wilson-Dirac operator is
 \begin{equation}
D_{ee} + D_{oo} = (4+m_0) \exp\left[\frac{\csw}{4+m_0}\frac{i}{4}\sigma_{\mu\nu}\widehat{F}_{\mu\nu}\right].
 \end{equation}
Here $\psi = (\mathfrak{u}, \mathfrak{d})^T$, $\beta= 4/g_0^{2}$, $m_0$ is the bare fermion mass, $\widehat{F}_{\mu\nu}$ is the lattice gauge field strength, $\sigma_{\mu\nu} = \frac{i}{2}[\gamma_\mu, \gamma_\nu]$ and $\csw$ is the Sheikholeslami-Wohlert parameter. $D$ is diagonal in flavour space.

We consider box sizes of volume $V=L^3 \times T$, and we write quantities in lattice units. The boundary conditions for the fermion fields and the gauge field are periodic in all four directions. The numerical simulations are performed using an improved version of the HiRep code first described in~\cite{Del_Debbio_2010}. We also make use of the Hasenbusch determinant factorisation to speed up our simulations to explore the chiral regime of the gauge theory we are focusing on.

As previously mentioned, we label states by $SU(2)$ isospin, a subgroup of the flavour symmetry group $Sp(4)$, and in this work we restrict to just isovector states.
The interpolating fields considered are:
\begin{equation}
  \label{eq:interpolating_field}
O_{\Gamma} =  \bar{\mathfrak{u}} \Gamma\mathfrak{u}(x) - \bar{\mathfrak{d}} \Gamma\mathfrak{d}(x),\quad\textrm{with}\quad \Gamma = \gamma_5, \gamma_\mu\,.
\end{equation}

We will refer to states which transform as pseudoscalars under the Lorentz group with a subscript $\mathrm{PS}$, and to states which transform as vectors under the Lorentz group with a subscript $\mathrm{V}$.

 \section{Simulation Results} 
 \subsection{Non-perturbative tuning of $\csw$  using Schr\"{o}dinger Functional Simulations}
  
\begin{figure}
\centering
\includegraphics[width=1.0\textwidth]{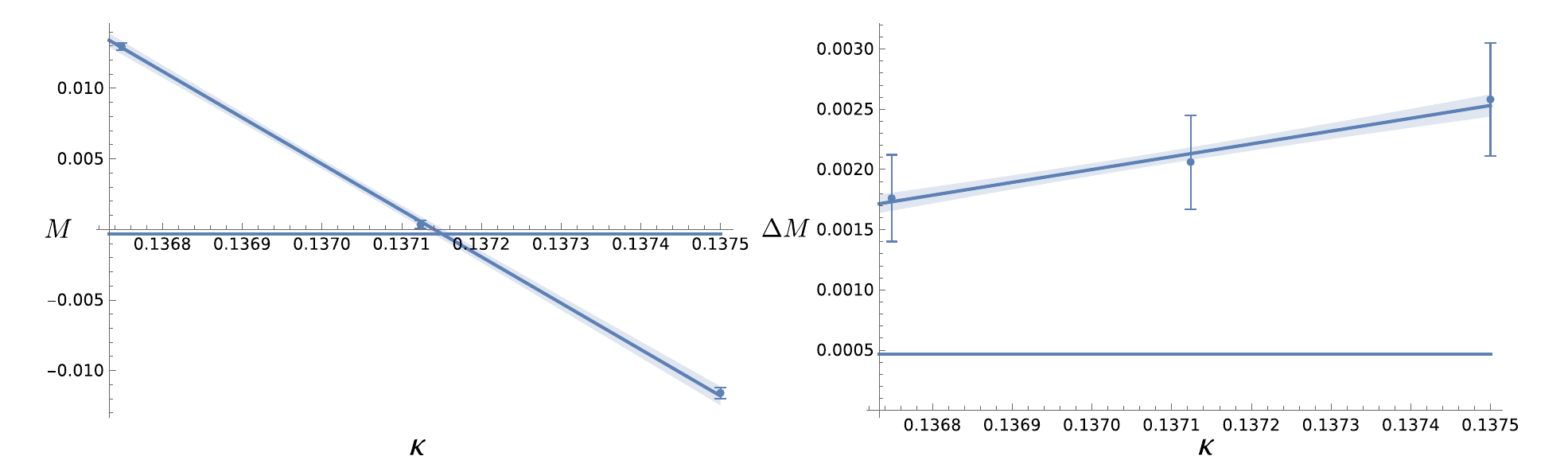}
\caption{Results from the Schr\"{o}dinger functional simulations used in the tuning.\\ \textit{Left:} The unrenormalised PCAC mass as a function of the hopping parameter $\kappa = \frac{1}{2am_0 + 8}$, with a linear interpolation to find $\kappa_{\mathrm{crit}}$ at fixed $\csw$.\\ \textit{Right:} $\Delta M$ as a function of $\kappa$. The horizontal line represents the perturbation theory result. When $\csw$ is tuned, $\Delta M$ at $\kappa_{\mathrm{crit}}$ will be equal to its value defined in perturbation theory.}
\label{fig:csw_tuning}
\end{figure}

\begin{figure}
\centering
\includegraphics[width=0.63\textwidth]{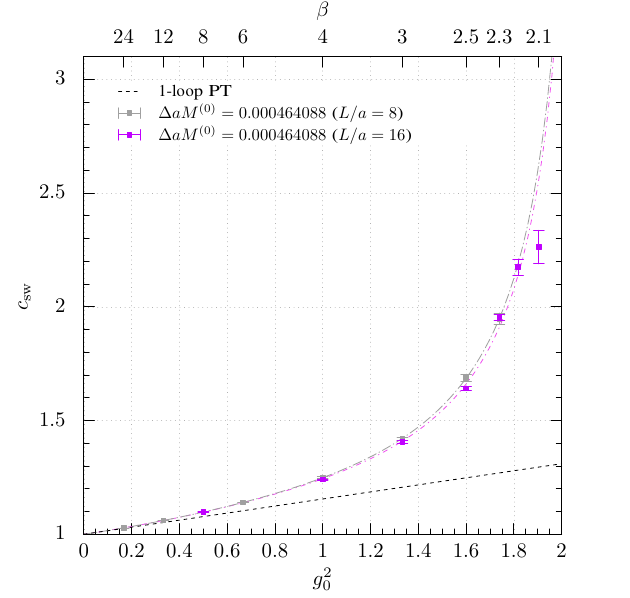}
\caption{The tuned value of $\csw$ as a function of the gauge coupling for two different volumes and compared to the $1$-loop PT result. The fit is a Pad\'{e} approximant where the first two terms are matched with the 1-loop PT result.}
\label{fig:csw}
\end{figure}

The parameter $\csw$ must be tuned non-perturbatively, which is accomplished by performing simulations with Schr\"{o}dinger functional boundary conditions following the well-established procedure~\cite{L_scher_1997}. For a fixed value of $\beta$, first the critical mass $m^{\mathrm{crit}}_0(\beta,\csw)$ is determined. This is achieved by imposing that the unrenormalised PCAC mass, denoted $M$, matches its tree level value. To determine the value of $\csw(\beta)$ which provides $O(a)$ improvement, a second matching condition is required. The condition requires that a linear combination $\Delta M(\beta, m^{\mathrm{crit}},\csw)$ defined through Ward identities vanishes up to tree level corrections.
 The tuning procedure is illustrated in figure~\ref{fig:csw_tuning}, and the results are displayed in figure~\ref{fig:csw}.

We ran simulations with a volume of $16^4$ for various values of $\beta$, $m_0$ and $\csw$ in order to tune $\csw$. We also ran simulations with a volume of $8^4$ to investigate finite volume effects. As can be seen in figure~\ref{fig:csw} our results match the 1-loop perturbation theory results shown by the dashed line in the small coupling regime.
In the large coupling region we observe large deviations from the perturbation theory result. We also observe possible signs of non-monotonicity for $\beta < 2.2$. The data are well described by a polynomial fit in the region $\beta > 2.2$ as shown by the dash-dot lines.


\subsection{Scale Setting and Topological Charge}
\begin{figure}
\begin{subfigure}[t]{0.5\textwidth}
\centering
\includegraphics[width=1.0\textwidth]{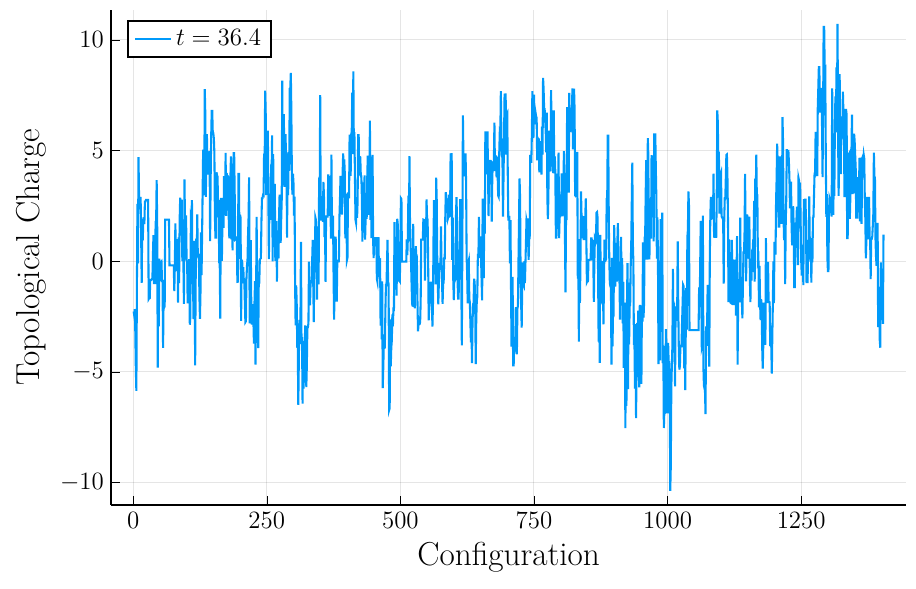}
\caption{Plot of the topological charge history measured at the Wilson flow time $t = w_0^2$.}
\label{fig:tc}
\end{subfigure}
\begin{subfigure}[t]{0.5\textwidth}
\centering
\includegraphics[width=1.0\textwidth]{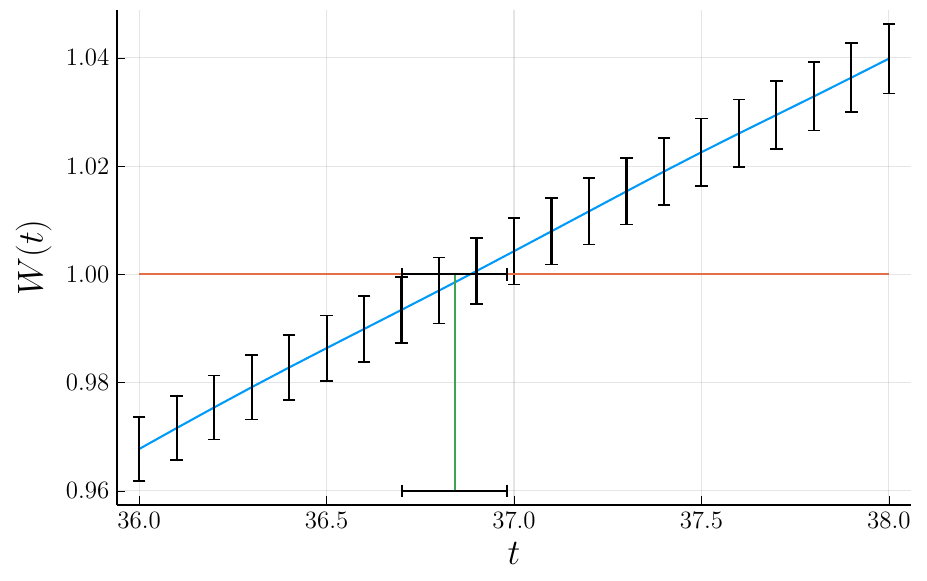}
\caption{The interpolation determining $w_0$.}
\label{fig:w}
\end{subfigure}
\caption{Results from a simulation at the finest lattice spacing. The parameters of this simulation are $\beta = 2.3$. $V = 48^4$, $m = -0.2987$.}
\end{figure}

We set the scale using the Wilson gauge flow. We define the reference scale $w_0$ for a given ensemble by $W(w_0^2) = 1.0$, where $W(t) = t\frac{d}{dt}\left\{t^2\langle E(t) \rangle_{\mathrm{sym}} \right\}$ following the definitions in \cite{w0paper}. We use symmetric finite differences in the definition of $E(t)$ and $W(t)$. In physical units, observables such as masses obtained from the simulation are dimensionless quantities $am$ carrying a factor of the lattice spacing, and the measured $a^{-1}w_0$ carries a factor of the inverse lattice spacing, so $a^{-1}w_0 \times am = w_0m$ is a quantity independent of the lattice spacing. This allows masses calculated on ensembles with different lattice spacings to be compared. In practice we average $W(t)$ over configurations, and then interpolate to solve for $w_0$. In order to estimate the error while managing autocorrelations in the data, the whole analysis is performed in a two-layer bootstrap to estimate the error and the error on the error for a range of bin sizes. Then the optimum bin size is chosen by plotting the error on the error against the bin size and choosing the bin size for which the error plateaus. In smearing the gauge field using the Wilson flow, it is important to monitor the smearing radius $c = \sqrt8 w_0$ to make sure that we do not oversmear and introduce unwanted effects due to the periodic boundary conditions. For all our ensembles we have $c < 0.4L$. We monitor the topological charge history measured at the reference Wilson flow time $t = w_0^2$ in order to make sure the simulations are not frozen. Figure~\ref{fig:tc} is a plot showing the topological charge history for an ensemble at our finest lattice spacing $\beta = 2.3$, and while the autocorrelation time is large, the topological charge is not frozen.
Here we work with $w_0$ at finite PCAC mass, and at this stage do not extrapolate to the chiral limit.

\subsection{Spectroscopy}
\begin{figure}[h]
\centering
\includegraphics[width=0.7\textwidth]{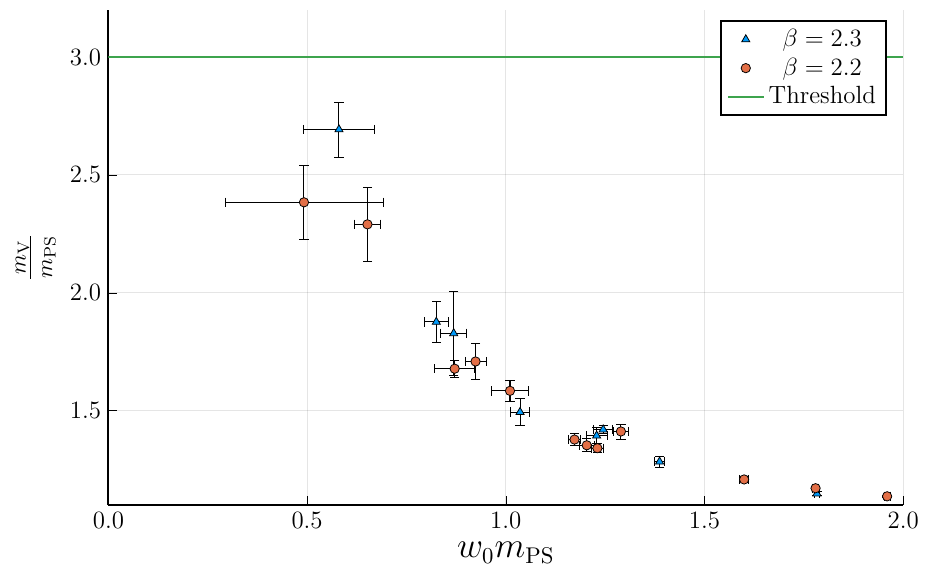}
\caption{The mass of the lightest vector isotriplet state to the lightest pseudoscalar state as a function of the lightest pseudoscalar state.}
\label{fig:mvmpsb22}
\end{figure}

We only consider ensembles with $Lm_{\mathrm{PS}} > 5$ to ensure that there is no contamination from finite volume effects. The correlators $f_\Gamma[t]$ are folded according to the periodicity in lattice time $t$, where $\Gamma$ stands for some combination of $\gamma$ matrices. The effective mass $m_\Gamma^{\mathrm{eff}}[t]$ is then computed by implicitly solving

\begin{equation} \frac{f_\Gamma[t-1]}{f_\Gamma[t]} = \frac{e^{-(T - (t-1))m_\Gamma^\mathrm{eff}[t]} + e^{-(t-1)m_\Gamma^\mathrm{eff}[t]}}{e^{-(T - t)m_\Gamma^\mathrm{eff}[t]} + e^{-tm_\Gamma^\mathrm{eff}[t]}}.
\end{equation}
In order to extract the ground state energy, a constant is then fitted to $m_\Gamma^{\mathrm{eff}}[t]$ in the plateau region. To determine the error, the same binning double bootstrap procedure is performed as for the scale setting. 
In this preliminary work we use ensembles with $\beta = 2.2$ and $\beta = 2.3$. Figure~\ref{fig:mvmpsb22} shows the dependence of $\frac{m_{\mathrm{V}}}{m_{\mathrm{PS}}}$ against $w_0m_{\mathrm{{PS}}}$ at the two different lattice spacings, which provides confidence that the results are not strongly affected by finite $a$ effects. In this new setup at $\beta = 2.3$, $\mv$ has reached the maximum value for which the singlet state is known to be stable.
\begin{figure}[h]
\centering
\includegraphics[width=0.7\textwidth]{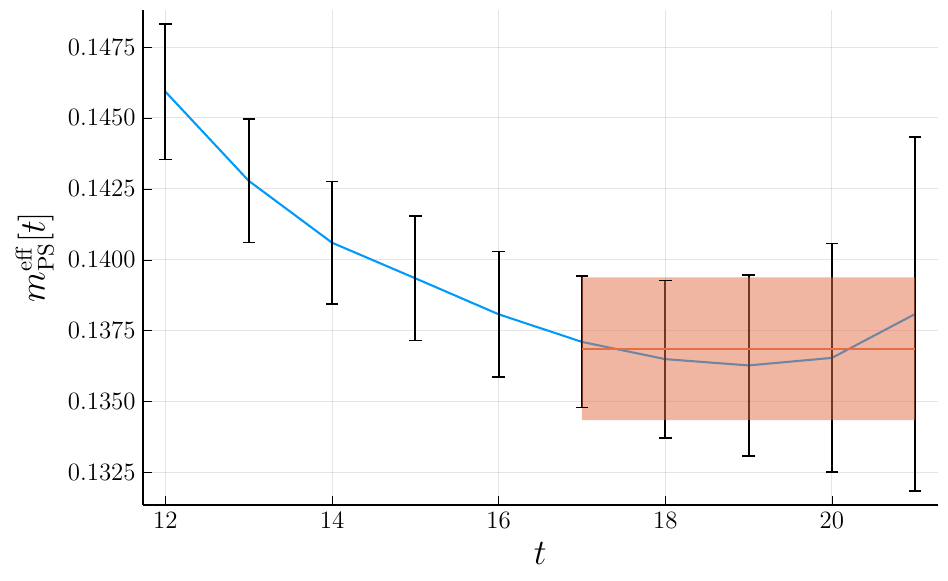}
\caption{An effective mass plot for the pseudoscalar channel showing the effective mass plateau and the fit, again at $\beta = 2.3$, $V = 48^4$, $m = -0.2987$. The points are connected to guide the eye.}
\end{figure}

\section{Conclusion}
Preliminary results have been presented using the new exponential clover action for $SU(2)$ with two fundamental flavours, a minimal model for the composite Higgs sector in isolation. We have obtained tuned values for $\csw$ in the exponential clover action, giving $O(a)$ improvement for any $\beta \geq 2.2$. Physics simulations were then performed using the tuned action, for which the finite volume effects and lattice discretisation error are well under control, and we are now simulating the region of parameter space where $\frac{\mv}{\mps} > 2.5$ for two different lattice spacings. Our long term goal is to compute scattering amplitudes in the singlet channel and constrain the phenomenology of the Higgs boson at the LHC. To this end we require excellent control of the systematic errors, particularly the discretisation error. This is the reason for using the exponential clover action. We are now performing simulations at two different lattice spacings at the scale where we expect the singlet resonance to be unstable. The next simulations will provide insight into the range of validity of the underlying effective field theory that describes the composite Higgs sector in isolation of the Standard Model. We will then be able to provide non-perturbative estimates of the low energy constants of the theory which can be used to constrain the phenomenology of composite Higgs models. The next steps are to perform a continuum limit extrapolation of the spectroscopic quantities, and then to perform scattering calculations in this new setup with these and more chiral runs.
\bibliographystyle{JHEP}
\bibliography{bib}

\end{document}